\documentstyle[prl,aps,epsf,twocolumn]{revtex}

\catcode`\@=11
\def\simle{\mathrel{\mathpalette\@versim<}}   
\def\simge{\mathrel{\mathpalette\@versim>}}   
\def\@versim#1#2{\lower2.5pt\vbox{\baselineskip0pt \lineskip-.5pt
   \ialign{$\m@th#1\hfil##\hfil$\crcr#2\crcr\sim\crcr}}}

\begin{document}

\draft
\twocolumn[\hsize\textwidth\columnwidth\hsize\csname @twocolumnfalse\endcsname

\title{
One-Dimensional Confinement and Enhanced \\
Jahn-Teller Instability in LaVO$_3$
}
\author{
Yukitoshi Motome$^1$, Hitoshi Seo$^{2,3}$,
Zhong Fang$^1$, and Naoto Nagaosa$^{1,2,4}$
}
\address{
$^{1}$Tokura Spin SuperStructure Project (SSS), ERATO,
Japan Science and Technology Corporation (JST),
c/o National Institute of Advanced Industrial Science and Technology (AIST),
Tsukuba Central 4, 1-1-1 Higashi, Tsukuba, Ibaraki 305-8562, Japan
}
\address{
$^{2}$Correlated Electron Research Center (CERC), AIST,
Tsukuba Central 4, 1-1-1 Higashi, Tsukuba, Ibaraki 305-8562, Japan
}
\address{
$^{3}$Domestic Research Fellow, JST,
4-1-8 Honcho, Kawaguchi, Saitama 332-0012, Japan
}
\address{
$^{4}$Department of Applied Physics, University of Tokyo,
7-3-1, Hongo, Bunkyo-ku, Tokyo 113-8656, Japan
}
\date{\today}

\maketitle

\begin{abstract}
Ordering and quantum fluctuations of orbital 
degrees of freedom are studied theoretically for LaVO$_3$ in 
spin-C-type antiferromagnetic state. 
The effective Hamiltonian for the orbital pseudospin shows
strong one-dimensional anisotropy 
due to the negative interference among various exchange processes. 
This significantly enhances the instability toward lattice distortions
for the realistic estimate of the Jahn-Teller coupling
by first-principle LDA+$U$ calculations,
instead of favoring the orbital singlet formation.
This explains well the experimental results on
the anisotropic optical spectra as well as 
the proximity of the two transition temperatures
for spin and orbital orderings. 
\end{abstract}

]

Orbital degrees of freedom are playing key 
roles in magnetic and charge transport properties of transition metal 
oxides \cite{tokuranagaosa}. Especially it has been recognized that,
in these strongly-correlated systems, spatial shapes of 
orbitals can give rise to an anisotropic electronic state even in 
the three-dimensional (3D) perovskite structure
\cite{Kugel1982,Imada1998}.
There the spin ordering (SO) and orbital ordering (OO) 
are determined self-consistently \cite{maezono}.

Perovskite vanadium oxides, AVO$_{3}$ (A is rare-earth element),
are typical $t_{2g}$ electron systems which show this interplay 
between orbital and spin degrees of freedom
\cite{Mahajan1992,Nguyen1995,Kawano1994,Ren2000,Noguchi2000,Blake2001}.
Both magnetic and orbital transition temperatures, 
$T_{\rm N}$ and $T_{\rm o}$, respectively, change systematically
according to the ionic radius of the A atom
\cite{MiyasakaPC},
which controls
the bandwidth through the tilting of VO$_6$ octahedra.
For smaller ionic radii (smaller bandwidth) such as A=Y,
$T_{\rm o}$ for OO of G-type (3D staggered) is much higher than
$T_{\rm N}$ for SO of C-type (rod-type)
\cite{Blake2001,MiyasakaPC}.
As the ionic radius increases (the bandwidth increases),
$T_{\rm o}$ decreases while $T_{\rm N}$ increases, and
finally they cross between A=Pr and Ce
\cite{MiyasakaPC}.
In LaVO$_3$, 
the SO occurs at $T_{\rm N} \simeq 143$K first, 
and at a few degrees below $T_{\rm N}$ 
the OO takes place
\cite{Miyasaka2000}.
A remarkable aspect here is its proximity of
$T_{\rm N}$ and $T_{\rm o}$,
which is also observed
for all the compounds with $T_{\rm N} > T_{\rm o}$, 
i.e., CeVO$_3$ 
\cite{MiyasakaPC}
and La$_{1-x}$Sr$_{x}$VO$_{3}$ ($x < 0.17$)
\cite{Miyasaka2000}.
Therefore, in LaVO$_{3}$, 
the magnetic correlation appears to develop primarily and to 
induce the orbital transition immediately 
once the SO sets in.

\begin{figure}
\epsfxsize=6.5cm
\centerline{\epsfbox{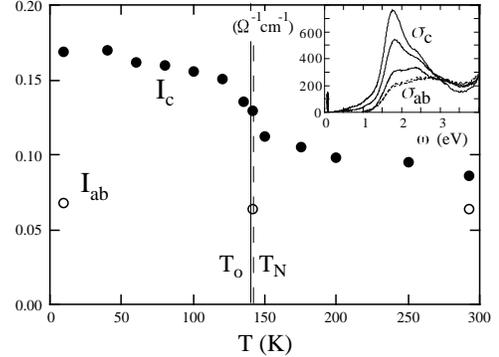}}
\caption{
Temperature dependence of the spectral weight in LaVO$_{3}$.
Filled and open circles represent the data along the $c$ axis 
$I_c$ and within the $ab$ plane $I_{ab}$, respectively.
The inset shows the optical conductivity.
The solid curves denote the data along the $c$ axis
for $T=10, 142, 293$K from top to bottom, respectively.
The data within the $ab$ plane are almost temperature independent 
which are shown by the dashed curves.
}
\label{fig:experiment}
\end{figure}

Another interesting aspect of LaVO$_3$ is the large anisotropy 
in the electronic state, which has recently been explored by
the optical spectra \cite{Miyasaka2002}.
Figure~\ref{fig:experiment} shows
the temperature dependence of the spectral weights,
$I_c$ along the $c$ direction and $I_{ab}$ within the $ab$ plane,
which is obtained from the data in ref.~\ref{Miyasaka2002}.
Here we define the spectral weight as an integration of
the optical conductivity up to the isosbestic (equal-absorption) 
point at $2.8$eV, namely,
$I_{\mu} \equiv \frac{2m_0}{\pi e^2 n} \int_0^{2.8{\rm eV}}
\sigma_{\mu}(\omega) d\omega$
($\mu = c$ or $ab$),
where $m_0$ and $n$ are the free electron mass and 
the density of V atoms, respectively.
The most striking feature is the temperature dependence.
$I_{c}$ grows rapidly below $T_{\rm N}$ while $I_{ab}$
is almost temperature independent.
Therefore the temperature dependence is 
almost 1D although the ratio $I_c/I_{ab} \sim 2$ is not so large. 

Compared with $e_g$ systems, the Jahn-Teller (JT)
coupling in $t_{2g}$ systems is expected to be weak, and 
the quantum fluctuation and/or the singlet formation
of the orbital degrees of freedom is a keen 
issue. Recently, Khaliullin {\it et al.} 
claimed that orbital singlet correlation
along the $c$ axis is the driving force to realize the 
ferromagnetic spin exchange in the C-type antiferromagnetic (AF) phase 
in LaVO$_3$ \cite{Khaliullin2001}.
On the other hand, the C-type SO state with the G-type OO 
has been obtained by the mean-field theory
\cite{Mizokawa1996} and the first-principle calculation
\cite{Sawada1996} which are justified for rather 
weakly-correlated cases and 
do not take account of quantum fluctuations seriously.
There, the AF interactions within the $ab$ plane 
concomitant with a single occupation of $xy$ orbitals
play a key role to stabilize the SO and OO.
Hence, these two pictures are quite different.
Since there are competing interactions 
with the orbital quantum nature,
such as JT coupling and 3D orbital exchange couplings,
it is highly nontrivial to what extent
the quantum fluctuations are important under the realistic situations,
nevertheless the quantitative study on this issue has been missing.

In this Letter, we study ordering and fluctuations
of orbital degrees of freedom
in the C-type AF phase in LaVO$_{3}$. 
An effective orbital model 
including the JT and the relativistic spin-orbit 
couplings is derived using the parameters obtained by 
first-principle LDA+$U$ calculations and the optical experiments.
We show that the model exhibits a strong 1D 
anisotropy which explains well the experimental results 
for the optical spectra and the proximity of $T_{\rm o}$ to $T_{\rm N}$.
It is concluded that the enhanced JT instability due to the 
1D confinement dominates the orbital singlet formation in LaVO$_3$.

Now we derive the effective orbital model. 
We start from the strong-coupling limit
of the Hubbard model with three-fold orbital degeneracy
for the $t_{2g}$ orbitals
\cite{Kugel1982}.
The system contains two $d$ electrons at each V atom 
which form the high-spin $S=1$ state due to the Hund's-rule coupling.
To focus on the orbital sector in this spin-orbital coupled Hamiltonian,
we assume the C-type SO.
This is a reasonable approximation because 
the spin $S=1$ has less quantum nature compared to $S=1/2$ 
and the C-type SO is obtained
by the mean-field calculation
\cite{Mizokawa1996} 
and the first-principle calculation
\cite{Sawada1996}
as mentioned above.
At the same time, we assume that 
the $xy$ orbital is singly occupied at each V atom, 
which drives the AF coupling in the $ab$ plane
\cite{Mizokawa1996,Sawada1996}.
Then the second electron goes to either $yz$ or $zx$ orbital. 
We assign a pseudospin state $\tau^{z} = \pm 1/2$ for the occupancy of 
these two orbitals
\cite{Khaliullin2001}.
Finally, our total Hamiltonian is written as
$
{\cal H} = {\cal H}_{\rm orb}^c + {\cal H}_{\rm orb}^{ab}
+ {\cal H}_{\rm JT} + {\cal H}_{LS},
$
each term of which is described below.

The orbital exchange term for the $c$ direction is
\begin{equation}
\label{eq:H_orb^c}
{\cal H}_{\rm orb}^c = \sum_{\langle ij \rangle} 
J_c ({\vec \tau}_i \cdot {\vec \tau}_j - 1/4),
\end{equation}
while that within the $ab$ plane is
\begin{equation}
\label{eq:H_orb^ab}
{\cal H}_{\rm orb}^{ab} = \sum_{\langle ij \rangle} 
\big( J_{ab}^- \tau_i^z \tau_j^z
- J_{ab}^+/4 - J_{ab}^{xy}/4 \big).
\end{equation}
Here the summations are taken for nearest-neighbor pairs.
The coupling constants are given by
\begin{eqnarray}
J_c &=& 4t^2/(U'-J_{\rm H}),
\label{eq:J_c}
\\
J_{ab}^{\pm} &=&
\frac{2t^2}{3(U'-J_{\rm H})} + \frac{4t^2}{3(U'+2J_{\rm H})}
\pm \frac{t^2}{U+2J_{\rm H}} \pm \frac{t^2}{U},
\label{eq:J_ab}
\\
J_{ab}^{xy} &=&
4t^2/(U+2J_{\rm H}) + 4t^2/U,
\label{eq:J_ab^xy}
\end{eqnarray}
where $U, U'$ and $J_{\rm H}$ are the intra-orbital, the inter-orbital
Coulomb interaction and the Hund's-rule coupling, respectively.
Neglecting the small tilting of VO$_6$ octahedra in LaVO$_3$,
the transfer integral $t$ is taken to be diagonal which
strongly depends on the direction, for instance,
$t_{ij}^{yz} = t_{ij}^{zx} = t$ and otherwise zero
in the $c$ direction.
From the analysis in ref.~\ref{Miyasaka2002},
the parameters are estimated as $U \simeq 2.25$eV, $U' \simeq 1.93$eV
and $J_{\rm H} \simeq 0.16$eV.
The transfer integral $t$ is set to be $0.12$eV
based on the estimate of the bandwidth $\simeq 1$eV
in first-principle calculations
\cite{Sawada1996}.
Then the orbital exchange interaction along the $c$ axis
is estimated as $J_c \simeq 33$meV, while $J_{ab}^- \simeq 2$meV.
We set $J_c = 1$ as an energy unit in the following calculations.

Here we point out two important features in 
eqs.~(\ref{eq:H_orb^c}) and (\ref{eq:H_orb^ab}).
One is that the exchange in the $c$ direction is
Heisenberg-type while that within the $ab$ plane is Ising-type.
From this, one might expect a strong quantum fluctuation
in the $c$ direction as pointed out 
in ref.~\ref{Khaliullin2001}.
However, this quantum nature becomes relevant only 
when the JT coupling is negligibly small, and
this is not the case in LaVO$_3$ as discussed in the following.
The other important feature is the large 1D anisotropy
in the orbital exchange couplings.
Note that the negative interference among
different perturbation processes occurs
in the in-plane coupling $J_{ab}^-$ in eq.~(\ref{eq:J_ab}),
which results in the ratio of the exchange couplings 
$J_c / J_{ab}^- \sim 17$.

\begin{figure}
\epsfxsize=6.5cm
\centerline{\epsfbox{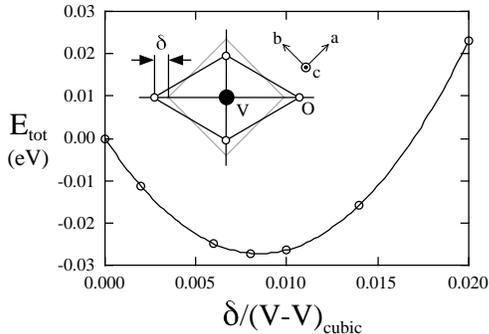}}
\caption{
Total energy per V atom as a function of
the Jahn-Teller distortion obtained 
by the LDA+$U$ calculation.
The inset shows the tetragonal distortion of VO$_6$ octahedra
considered here. 
The curve is a fit by a quadratic polynomial.
}
\label{fig:JT}
\end{figure}

The JT coupling in the subspace of $\tau$ is given by
\begin{equation}
\label{eq:H_JT}
{\cal H}_{\rm JT} = \sum_{i} g Q_{i} \tau_{i}^{z} + \frac{1}{2} \sum_{i} 
Q_{i}^{2},
\end{equation}
where $Q_{i}$ is the JT phonon coordinate at site $i$.
We neglect the kinetic energy of phonons,
namely, regard $Q_{i}$ as a classical variable. 
It is crucial to estimate the coupling constant $g$, and we have done 
the following first-principle calculation
\cite{Fang2002}.
Assuming the tetragonal symmetry, we calculate the total energy
as a function of the JT distortion as shown in Fig.~\ref{fig:JT}.
This gives the JT stabilization energy $\sim 27$meV, 
which approximately corresponds to $E_{\rm JT} \equiv g^2/8$ in our model.
Thus we obtain the estimate $E_{\rm JT} \simeq 0.8 J_c$ ($g \simeq 2.6$),
which is appreciable and cannot be neglected as in ref.~\ref{Khaliullin2001}.

The last term is the relativistic spin-orbit coupling,
which may be important in the $t_{2g}$ systems.
We obtain the effective Hamiltonian by projecting 
the original form ${\cal H}_{LS} = \sum_{i} 
\lambda {\vec L}_{i} \cdot {\vec S}_{i}$
to the subspace of $\tau$
by using the experimental fact that the spins lie within the $ab$ plane
\cite{Zubkov1973}.
Here ${\vec L}_i$ is the orbital angular momentum.
Since $L_{x}$ has matrix elements between $xy$ and $(yz,zx)$
in this case, the spin-orbit interaction is represented by
$
{\cal H}_{LS} = \sum_{i} \zeta \tau^{z}_{i},
$
where $\zeta \equiv \lambda^2/\Delta$ and
$\Delta$ is the energy separation between $xy$ and
($yz,zx$) orbital levels.
This indicates that the 
spin-orbit coupling corresponds to the pseudo magnetic field 
along the $z$ direction.
Using $\lambda \sim 20$meV in V atom and the estimate of
$\Delta \sim 1$eV in the band calculation
\cite{Sawada1996},
we estimate $\zeta \sim 0.4$meV$\sim 0.01 J_c$.
This is small enough to be neglected in the following calculations.

\begin{figure}
\epsfxsize=7cm
\centerline{\epsfbox{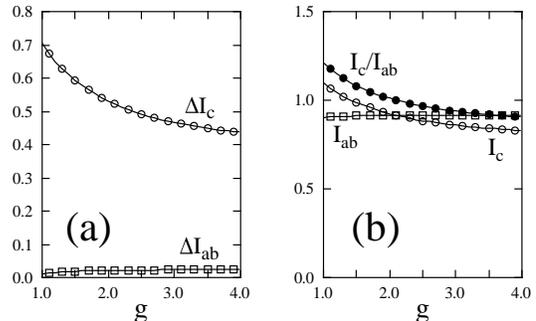}}
\caption{
(a) Contributions depending on temperature and
(b) total weights in the ground state for the spectral weight
as a function of the Jahn-Teller coupling $g$.
}
\label{fig:Lanzcos}
\end{figure}

First we discuss the anisotropic temperature dependence 
in the spectral weight.
The spectral weight is generally given 
by the kinetic energy of the system in the ground state $K$ as
$I = -\pi K / 4N$
where $N$ is the system size and we set $e^2=1$
\cite{Maldague1977}.
In the strong-coupling limit, this is calculated 
by the exchange energy as shown below.
Let us consider the Heisenberg model
for the strong-coupling limit of the Hubbard model at half filling.
The kinetic energy in the Hubbard model is calculated as
$K = t \partial \langle {\cal H} \rangle
/ \partial t = \partial \langle {\cal H} \rangle / \partial \ln t$.
Here the bracket denotes the expectation value in the ground state.
Since the exchange coupling $J$ in the Heisenberg model is proportional
to $t^2$, we obtain $I = - (\pi/4N) \times 2 \partial \langle {\cal H} \rangle 
/ \partial \ln J = -\pi \langle {\cal H}_{\rm Heis} \rangle / 2N$ where
${\cal H}_{\rm Heis}$ is the Heisenberg Hamiltonian.
Using this formula,
we calculate the spectral weight for the present model 
by the exact-diagonalization in the ground state
for $4 \times 4$-site lattice embedded in the $ac$ plane.
In Fig.~\ref{fig:Lanzcos} (a),
the obtained values of 
$\Delta I_c \equiv -\pi J_c \sum
\langle \vec{\tau}_i \cdot \vec{\tau}_j \rangle / 2N$ and
$\Delta I_{ab} \equiv -\pi J_{ab}^- \sum
\langle \tau_i^z \tau_j^z \rangle / 2N$ 
are plotted as a function of the JT coupling $g$.
Note that these exchange correlations correspond to 
the enhancements of the spectral weights 
from the high-temperature limit to the ground state.
The results indicate that,
the contribution in the $c$ axis 
is much larger than that in the $ab$ plane.
The ratio becomes larger than $20$ 
for the realistic value of $g \simeq 2.6$.
This explains well the anisotropic temperature dependence of
the spectral weight in Fig.~\ref{fig:experiment}.

We also discuss the magnitude of the total spectral weight 
in the ground state.
Figure~\ref{fig:Lanzcos} (b) shows the calculated spectral weights
$I_{\mu} = -\pi \langle {\cal H}_{\rm orb}^{\mu} \rangle / 2N$ 
which include the contributions from the constants 
in eqs.~(\ref{eq:H_orb^c}) and (\ref{eq:H_orb^ab}).
As shown in Fig.~\ref{fig:Lanzcos} (b),
the ratio $I_c/I_{ab}$ is almost $1$
for the realistic value of $g$, which is much smaller than that of 
the temperature dependent part in Fig.~\ref{fig:Lanzcos} (a).
This ratio is comparable but smaller than  
the experimental result $I_c/I_{ab} \sim 2$ in Fig.~\ref{fig:experiment}.
This might be due to the facts that (i) the cut-off energy $2.8$eV for the 
integration is not large enough to take all the contribution of $I_{ab}$,
and (ii) the transfer integrals 
for the different orbitals are different because they depend on 
the energy difference $\Delta_{dp}$ between the $d$ and
the oxygen $p$ orbitals as the energy denominator.
Since the huge anisotropy in temperature dependences
is reproduced with the moderate anisotropy in the total weights,
we believe that the orbital 1D confinement in our model 
plays a major role in the anisotropic electronic state
in this material.

\begin{figure}
\epsfxsize=6.5cm
\centerline{\epsfbox{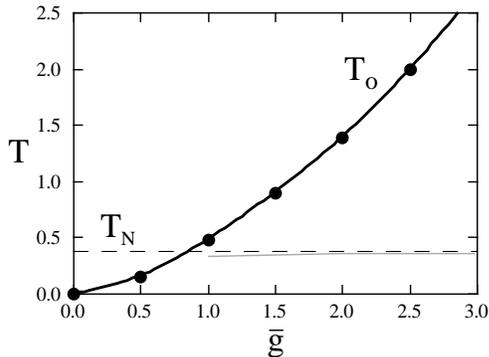}}
\caption{
Transition temperature of the orbital ordering
under the 1D confinement by the C-type AF spin order.
For comparison, the experimental value of $T_{\rm N}$
in LaVO$_3$ is shown by the dashed line. 
The gray curve shows the expected $T_{\rm o}^{\rm exp}$.
See the text for details.
}
\label{fig:phasediagram}
\end{figure}

We now turn to the discussion on the transition temperatures.
In order to obtain the phase diagram at finite temperatures,
we apply the transfer matrix method
\cite{tm} combined with the unrestricted Hartree-Fock approximation 
for the interchain coupling $J^-_{ab}$ \cite{Scalapino1975}.
We start from a configuration of $\{Q_i\}$
and calculate the expectation value $\langle \tau_i^z \rangle$ numerically.
Then, from the self-consistent equation $Q_i = - g \langle \tau_i^z \rangle$
which is obtained as the energy minimization for $Q_i$,
we define a new configuration of $\{Q_i\}$.
We repeat this procedure until $\{Q_i\}$ is optimized and
the energy is minimized.
In this treatment, $J^-_{ab}$ is effectively absorbed into 
the JT coupling $g$ as
$\sum_i (g Q_i - 2J_{ab}^- \langle \tau_i^z \rangle ) \tau_i^z
= - \sum_i \bar{g}^2 \langle \tau_i^z \rangle \tau_i^z$
where $\bar{g}^2 \equiv g^2 + 2J_{ab}^-$.
The obtained transition temperature for 
the orbital/lattice ordering transition
is plotted as a function of $\bar{g}$
in Fig.~\ref{fig:phasediagram}.
We note that for large $g$, the transition temperature
diverges as $T_{\rm o} \sim g^2/4$
which corresponds to the JT energy gap.
However, this is an artifact of the mean-field-type treatment, 
and in reality
$T_{\rm o}$ should stay at a constant of the order of $J_c$
since the coupling by $J_c$ between the neighboring sites
determines $T_{\rm o}$ in the limit of large $g$.

Considering the realistic value of $J_c \simeq 33$meV and
$E_{\rm JT} \simeq 27$meV ($g \simeq \bar{g} \simeq 2.6$) 
for LaVO$_3$, 
the orbital transition temperature is estimated as
$T_{\rm o} \sim 800$K
which is much higher than the observed $T_{\rm N} \simeq 143$K
as indicated in Fig.~\ref{fig:phasediagram}. 
One might think that 
this contradicts with the experimental fact 
$T_{\rm N} > T_{\rm o}^{\rm exp}$.  
However the phase diagram is obtained 
{\it assuming} the C-type SO, which induces the 1D confinement
of the orbital degrees of freedom 
with the enhanced $J_c$.
Note that the disorder of the spins should reduce
the effective orbital exchange
as easily shown in the spin-orbital coupled Hamiltonian.
It is also well-known that 1D systems have an 
enhanced instability to lattice distortions 
compared with higher dimensions. 
Therefore, $T_{\rm o}^{\rm 1D} = T_{\rm o}$ under 
this 1D orbital confinement can be higher than
$T_{\rm o}^{\rm 3D}$ without the SO
when the JT coupling governs the OO transitions.
In real materials with $T_{\rm o} > T_{\rm N}$, 
$T_{\rm o}^{\rm 3D}$ decreases as the bandwidth increases,
which indicates the relevance of the JT coupling.
(If the 3D orbital exchange couplings dominate,
$T_{\rm o}^{\rm 3D}$ should increase as $T_{\rm N}$ does.)
Then, when the inequality 
$T_{\rm o}^{\rm 3D} < T_{\rm N} < T_{\rm o}^{\rm 1D}$
is satisfied, the OO transition with the JT lattice distortion
should take place as soon as the SO grows and
induces the 1D confinement in the orbital channel.
In this scenario, 
comparing $T_{\rm o}^{\rm 1D}$ with $T_{\rm N}$,
we can estimate the lower bound for the value of $g$
to realize this proximity of the transition temperatures.
Our estimation for this lower bound from Fig.~\ref{fig:phasediagram}
is $g \sim 1$ 
which is consistent with the estimate in Fig.~\ref{fig:JT}.

Let us discuss this proximity of $T_{\rm N}$ and $T_{\rm o}^{\rm exp}$
by the Ginzburg-Landau type argument.
In this spin-orbital coupled system, we have the term
in which the SO parameter $M$ and the OO one $O$
are coupled as $\{a(T)-bM^2\}O^2$.
Here $a(T)$ is the coefficient for the second-order term
for OO without SO
which is given by $a(T)=a'(T-T_{\rm o}^{\rm 3D})$.
Our results indicate the relation
$
a(T)-bM_{\rm sat}^2 = a'(T-T_{\rm o}^{\rm 1D}),
$
where $M_{\rm sat}$ is the saturated magnetic moment.
$T_{\rm o}^{\rm exp}$ is given by solving the equation
$a(T_{\rm o}^{\rm exp})-bM^2( T_{\rm o}^{\rm exp}) = 0$.
Assuming $M(T) = M_{\rm sat} \sqrt{ (T_N-T)/T_N }$ for simplicity, 
the difference between $T_{\rm N}$ and $T_{\rm o}^{\rm exp}$
is given by 
$
\delta T / T_{\rm N} \equiv 
(T_{\rm N} - T_{\rm o}^{\rm exp}) / T_{\rm N}
= (T_{\rm N} - T_{\rm o}^{\rm 3D}) / (T_{\rm N} + \Delta T),
$
where $\Delta T = T_{\rm o}^{\rm 1D} - T_{\rm o}^{\rm 3D}$.
Considering the systematic changes of $T_{\rm N}$ and $T_{\rm o}$
for A-site ions
\cite{MiyasakaPC},
we expect that $T_{\rm o}^{\rm 3D}$ is slightly lower than
$T_{\rm N}$ in LaVO$_3$ and CeVO$_3$.
Assuming $T_{\rm o}^{\rm 3D} = 0.8 T_{\rm N}$, we
plot the expected $T_{\rm o}^{\rm exp}$
in Fig.~\ref{fig:phasediagram} as the gray curve.
For the realistic value of $g$, 
we have $T_{\rm o}^{\rm exp}$ quite close to $T_{\rm N}$
as observed in these compounds.

To summarize, we have investigated the role of orbitals 
to understand the electronic state in LaVO$_{3}$. 
We have derived the effective orbital model 
with strong one-dimensional anisotropy 
assuming the C-type spin ordering. 
We conclude that with the realistic Jahn-Teller coupling,
the orbital 1D confinement leads to an enhanced instability
toward lattice distortions suppressing the orbital quantum nature.
This gives a comprehensive description of 
the anisotropy in the optical spectra and
the proximity of the critical 
temperatures of magnetic and orbital transitions.

The authors appreciate Y. Tokura, S. Miyasaka, T. Arima, 
G. Khaliullin and B. Keimer for fruitful discussions.
Part of numerical computations are done using TITPACK ver.2
developed by H. Nishimori, who is also thanked.

\vspace*{-4mm}


\end{document}